\documentclass{DISproc}

\begin{document}
\title{Unbiased Polarised Parton Distribution Functions\\ and their Uncertainties}

\author{{\slshape Emanuele R. Nocera$^1$, Stefano Forte$^1$, Giovanni Ridolfi$^2$, Juan Rojo$^3$}\\[1ex]
$^1$Dipartimento di Fisica, Universit\`{a} di Milano and INFN, Sezione di Milano, \\Via Celoria 16 I-20133 Milano, Italy\\
$^2$Dipartimento di Fisica, Universit\`{a} di Genova and INFN, Sezione di Genova, \\Via Dodecaneso 33 I-16146 Genova, Italy\\ 
$^3$CERN, PH Department, TH Unit, CH-1211 Gen\`eve 23, Switzerland }

\contribID{273}

\doi  

\maketitle

\begin{abstract}
We present preliminary results on the determination
of spin-dependent, or polarised, Parton Distribution
Functions (PDFs) from all relevant inclusive polarised DIS data. The analysis
is performed within the NNPDF approach, which provides a faithful and
statistically sound representation of PDFs and their uncertainties.
We describe how the NNPDF methodology has been extended to the
polarised case, and compare our results with other recent polarised
parton sets. We show that polarised PDF uncertainties can
be sizeably underestimated in standard determinations, most notably for the gluon.
\end{abstract}

\noindent
The interest in spin-dependent, or polarised, Parton Distribution Functions (PDFs) of the nucleon
is mainly motivated by the desire to understand
its spin structure in terms of its quark and gluon parton substructure.
It largely originates
from the first EMC results~\cite{Ashman:1987hv}, originally interpreted
as an indication that quark and anti-quark
intrinsic angular momenta only contribute a small fraction of the full nucleon spin.
A faithful knowledge of polarised PDFs is 
also an essential ingredient 
for exploring QCD beyond the helicity-averaged case and for studying the phenomenology of 
spin-dependent processes. 

Polarised PDFs have been investigated with increasing precision in recent years.
On one hand, several experiments have contributed a large amount of data for a variety of processes, 
mainly inclusive polarised deep-inelastic scattering (DIS) but also proton-proton collisions 
and semi-inclusive reactions. 
On the theoretical side, the interest 
has been ultimately focussed on the global reconstruction of PDFs, together with their uncertainties.
At least four groups have constructed such polarised PDF sets recently: BB~\cite{Blumlein:2010rn}, 
AAC~\cite{Hirai:2008aj}, LSS~\cite{Leader:2010rb} and DSSV~\cite{deFlorian:2009vb}. 
These sets slightly differ in the choice of datasets, PDF parametrisation 
and details of the QCD analysis (such as the treatment of higher-twist corrections). 
Nevertheless, they are all
based on simple functional forms of the momentum fraction dependence
of the PDFs at the refence scale (typically, a power-like behavior
is assumed both at large and small momentum fraction)
and on the Hessian approach for the estimate of uncertainties.
Two main shortcomings are known to affect this methodology. 
The first one concerns how to propagate errors consistently from data 
to fitted parameters and then to observables: this is usually done by assuming Gaussian linear 
error propagation, which is not always adequate, in particular in those kinematical regions where few data are 
available.
The second one consists in assessing the theoretical bias introduced by a fixed functional parton
parametrisation. This is particularly delicate for polarised PDFs, owing to the quantity and the quality
of the data, which are respectively less abundant and less accurate than their unpolarised counterparts.

In order to overcome these difficulties,
in recent years the NNPDF collaboration has developed a new approach to parton fitting 
(see, for example,~\cite{DelDebbio:2007ee,Ball:2008by,Ball:2009mk,Ball:2010de,Ball:2011uy} and references therein). 
This new technique, designed to provide a faithful representation of PDFs and their uncertainties, is 
based on robust set of statistical tools, including Monte Carlo and Neural Network methods.

In the NNPDF approach, experimental data are sampled by generating an ensemble of Monte Carlo replicas
with data probability distribution; individual replicas are allowed to
fluctuate in such a way that
the mean value, standard deviation and correlation 
computed over Monte Carlo ensemble reproduce the experimental values, 
provided the sample is sufficiently large. Fitting 
an ensemble of parton distributions automatically propagates statistical fluctuations 
to the PDFs and then to observables. 
Hence, expectation values and uncertainties of PDFs (or of any observable) are obtained 
by considering their Monte Carlo integrals over the ensemble of replicas. 
Furthermore, in this approach
neural networks are used as unbiased interpolants for PDF parametrisation.
Since they provide functions depending on a large number of parameters, they are very flexible tools:
this flexibility allows one to reduce the bias associated to the choice of some fixed functional form.

\begin{wrapfigure}{r}{0.5\textwidth}
  \centering
  \includegraphics[width=0.35\textwidth,angle=270]{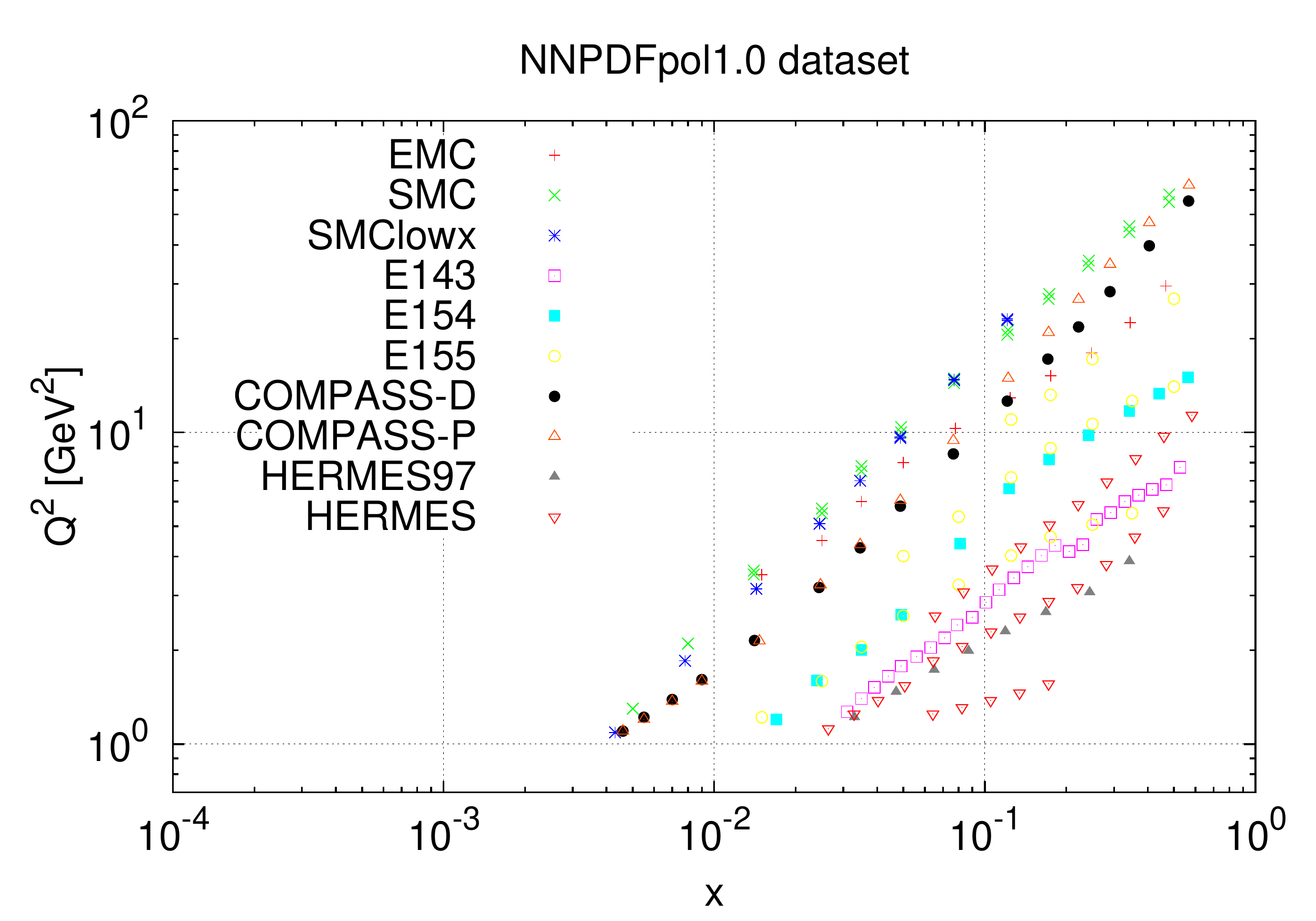}
  \caption{Experimental dataset after kinematic cuts for the \texttt{NNPDFpol1.0} analysis.}
  \label{Fig:kin}
\end{wrapfigure}

The NNPDF approach has been succesfully applied to the determination of unpolarised PDFs 
and these NNPDF sets are routinely used by Tevatron and LHC collaborations for data analysis 
and data-theory comparisons. We will present here
some preliminary results obtained by extending the NNPDF approach to the determination of a set of polarised PDFs. 
After illustrating the main features of our analysis, we will
compare our results to those obtained by other collaborations.
Specifically, we will see that the uncertainty on some polarised PDFs, most notably on the gluon PDF, 
are rather larger than previously estimated.       

\ \\ 
\noindent The first NNPDF analysis of polarised PDFs, \texttt{NNPDFpol1.0} henceforth, 
is based on a comprehensive set of polarised DIS data. 
We exclude from our analysis data points with $Q^2\leq Q^2_{\mbox{\scriptsize cut}}=
\unit{1}{\GeV^2}$, since below such energy scale perturbative QCD cannot be considered reliable. We also impose
$W^2 \geq W^2_{\mbox{\scriptsize cut}}=\unit{6.25}{\GeV^2}$ for the squared invariant mass
$W^2 = Q^2(1-x)/x$, according to the study presented in Ref.~\cite{Simolo:2006iw}. 
This choice removes the dependence of results on possible
dynamical higher-twist effects,
which we do not include even though we do include target-mass corrections.
The dataset used in the \texttt{NNPDFpol1.0} analysis is shown, after
kinematic cuts, in Fig.~\ref{Fig:kin}.   

The experimental data used in this fit do 
not allow a full separation of individual flavour and anti-flavour parton 
densities. Hence, we define, for each light flavour $q$, the net amount of quark-antiquark spin density
\begin{equation}
\Delta q(x,Q^2) = q^{\uparrow \uparrow}(x,Q^2)+\bar{q}^{\uparrow \uparrow}(x,Q^2)
                - q^{\uparrow \downarrow}(x,Q^2)+\bar{q}^{\uparrow \downarrow}(x,Q^2)
\mbox{ ,}
 \nonumber
\end{equation}
where the superscript $\uparrow\uparrow$ ($\uparrow\downarrow$) denotes that the parton spin is parallel 
(antiparallel) to the proton spin. We parametrise PDFs at the scale $Q_0^2=\unit{1}{\GeV^2}$ 
by choosing, besides the gluon density
$\Delta g(x,Q_0^2) \equiv g^{\uparrow\uparrow} - g^{\uparrow\downarrow}$, the following
three linear combinations of light quarks: the flavour-singlet
\begin{equation}
 \Delta\Sigma(x,Q_0^2) \equiv \Delta u(x,Q_0^2) + \Delta d(x,Q_0^2) + \Delta s(x,Q_0^2)
 \mbox{ ,}
 \nonumber
\end{equation}
the non-singlet triplet and the non-singlet octet
\begin{equation}
 \Delta T_3(x,Q_0^2) \equiv \Delta u(x,Q_0^2) - \Delta d(x,Q_0^2)
 \mbox{ ,}
 \ \ \ \ \ \ \
 \Delta T_8(x,Q_0^2) \equiv \Delta u(x,Q_0^2) + \Delta d(x,Q_0^2) - 2\Delta s(x,Q_0^2)
 \mbox{ .}
 \nonumber
\end{equation}
Each of these four combinations is parametrised by a neural network,
with a total number of $\mathcal{O}(200)$
parameters, to be compared to $\mathcal{O}(10 - 20)$ used in other
existing fits. 

A fast and accurate evaluation of polarised parton distributions, as required by the fitting, is achieved with the 
\texttt{FastKernel} method~\cite{Salam:2008qg}. 
The accuracy of polarised PDF evolution has been shown to be $\mathcal{O}(10^{-5})$
comparing with the \texttt{HOPPET} code.

Theoretical constraints are taken into account during the fitting procedure.
We have imposed positivity of physical cross-sections, which implies that
the polarised structure function $g_1$ is bounded by its unpolarised counterparts $F_1$,
so that $\left| g_1(x,Q^2)\right| \leq
F_1(x,Q^2)$~\cite{Altarelli:1998gn}. 
For consistency, the unpolarised structure functions have been 
computed from the recent NNPDF2.1 unpolarised PDF determination~\cite{Ball:2011uy}. 
We have also used SU(3) symmetry to relate the first moments
\begin{equation}
 a_3\equiv\int_0^1 dx \Delta T_3(x,Q_0^2)
 \mbox{ ,}
 \ \ \ \ \ \ \ \ \ \ 
 a_8\equiv\int_0^1 dx \Delta T_8(x,Q_0^2)
 \mbox{ ,}
 \label{Eq:sum_rules}
\end{equation}
to the determination of $a_3$ and $a_8$ from 
baryon decay constants (allowing for large uncertainties). 
We have also performed a fit 
in which we have relaxed the first constraint in Eq.~\ref{Eq:sum_rules} and we have considered $a_3$ as a
fit parameter, in which case we have found $a_3=1.21\pm 0.08$ to be compared with the global average from
experimental measurement of $\beta$-decay, $g_A=1.2701 \pm 0.0025$~\cite{Nakamura:2010zzi}. 
This result provides a consistency check of the fitting procedure
and validates the Bjorken sum rule with an accuracy of about $10\%$.

\begin{figure}[t]
 \centering
 \includegraphics[scale=0.35]{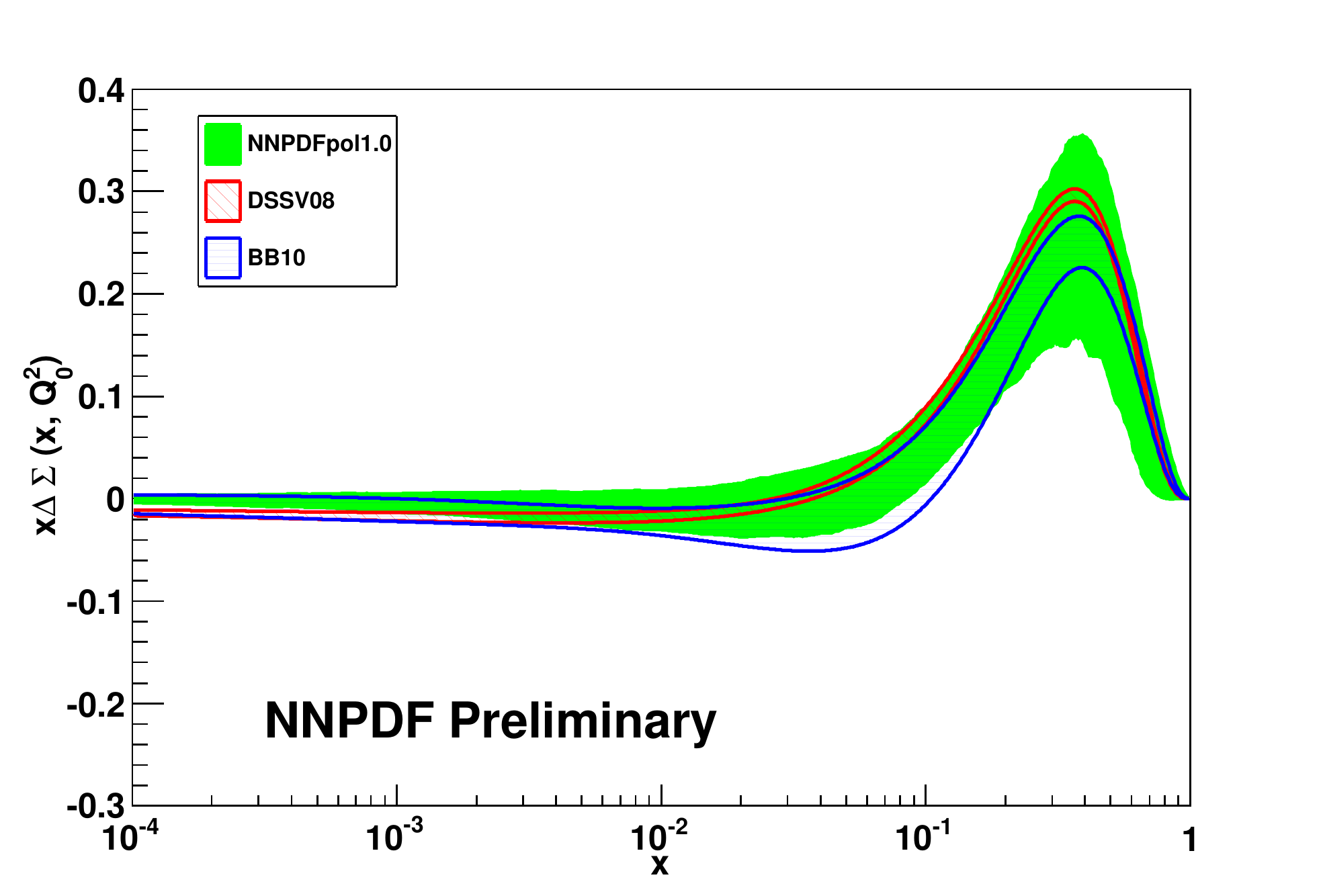}
 \includegraphics[scale=0.35]{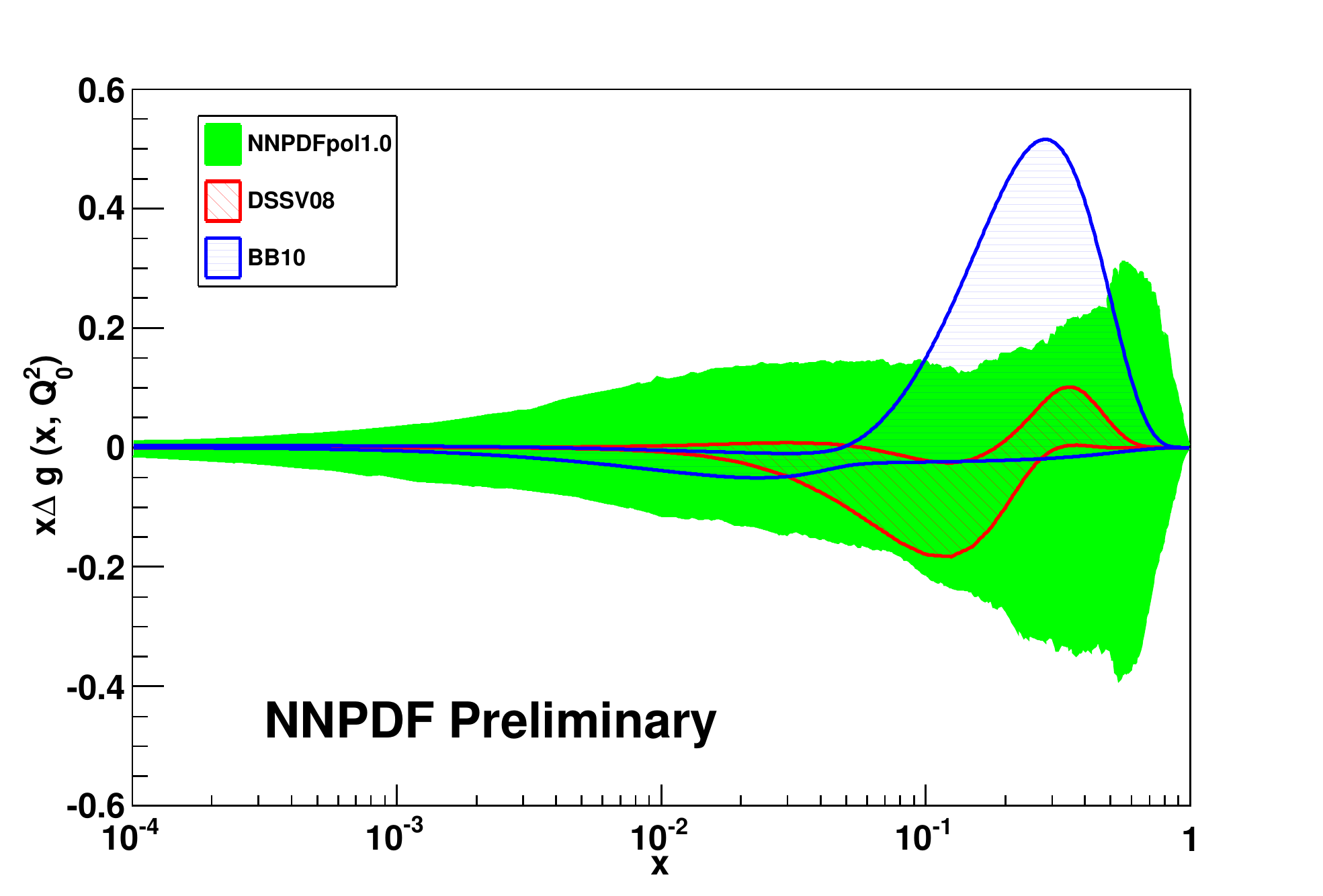}\\
 \includegraphics[scale=0.35]{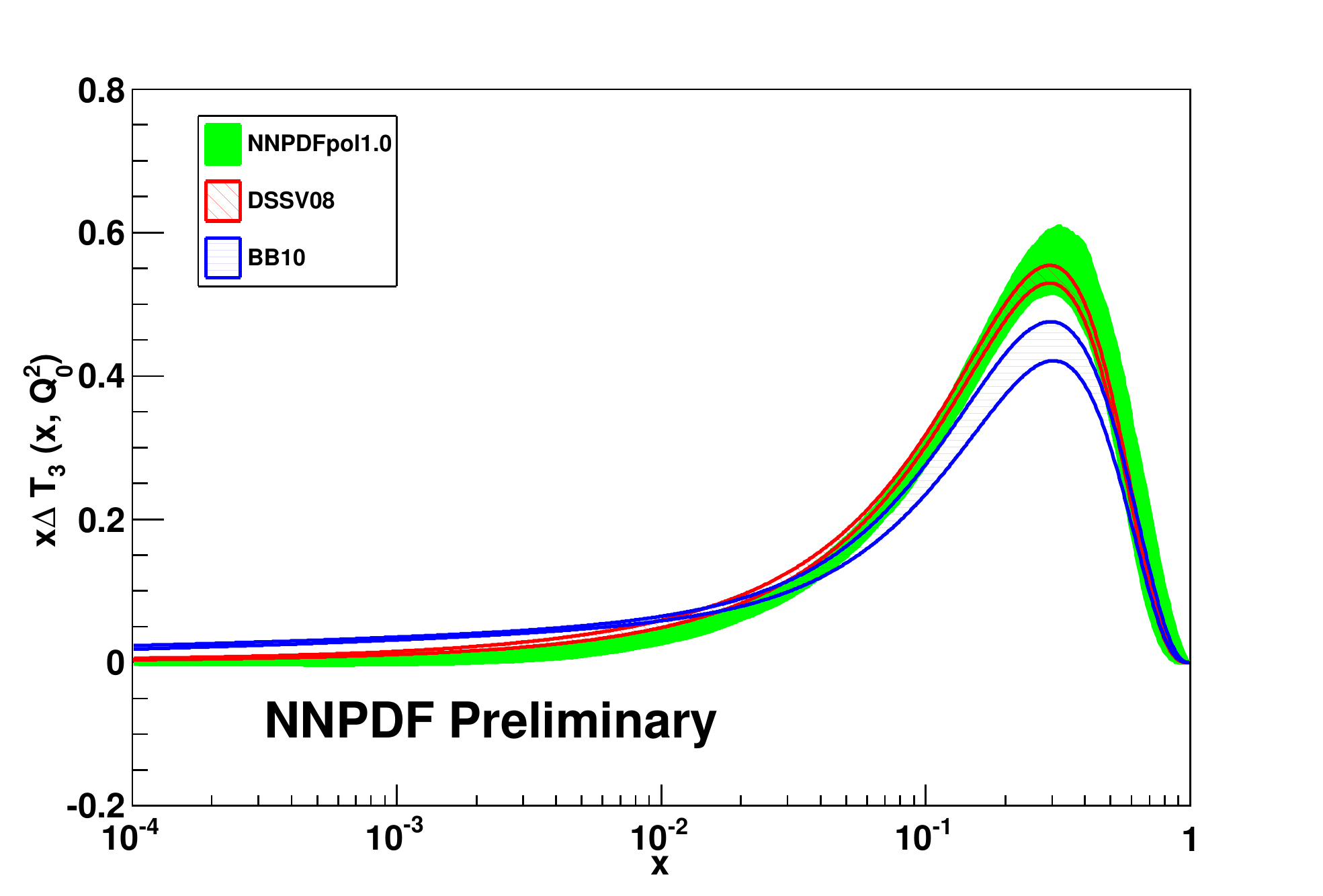}
 \includegraphics[scale=0.35]{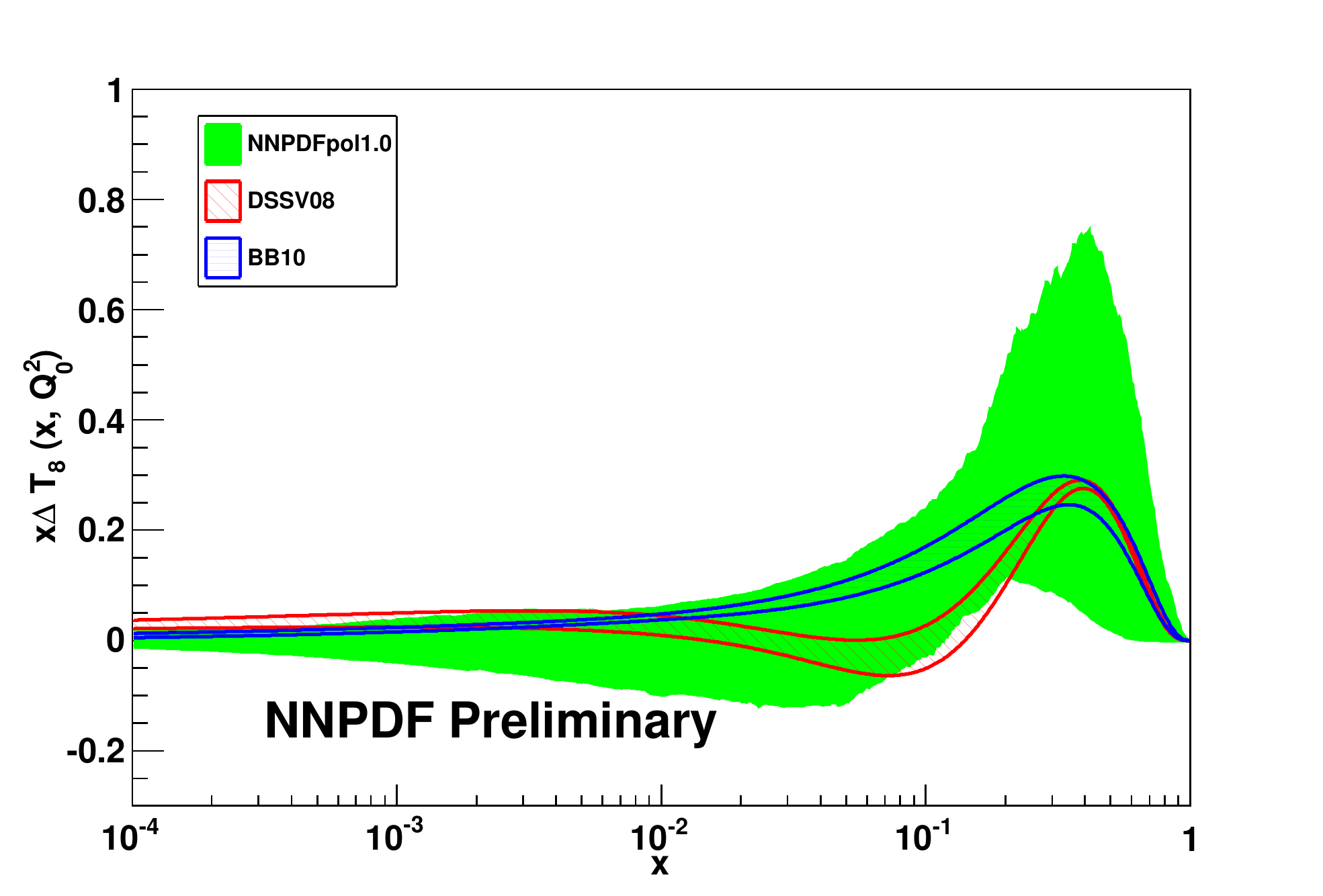}\\
 \caption{The \texttt{NNPDFpol1.0} parton set at the initial evolution scale $Q_0^2=\unit{1}{\GeV^2}$ compared to 
DSSV08~\cite{deFlorian:2009vb} and BB10~\cite{Blumlein:2010rn} determinations. Uncertainties on \texttt{NNPDFpol1.0}
parton distributions are computed at $68\%$ confidence level (see Ref.~\cite{Ball:2010de} for details).}
 \label{Fig:results}
\end{figure}

We show preliminary results for the \texttt{NNPDFpol1.0} set at initial scale
$Q_0^2=\unit{1}{\GeV^2}$ together with 
DSSV08~\cite{deFlorian:2009vb} and BB10~\cite{Blumlein:2010rn} determinations (Fig.~\ref{Fig:results}). 
In general, we can see that 
all PDFs show larger error bands than previously 
estimated, in particular at very small- or high-$x$ values, where no DIS data are available. 
This is especially the case for the polarised gluon PDF, which cannot be constrained by the available DIS data.
We also notice that, at least for the non-singlet triplet, the \texttt{NNPDFpol1.0}
analysis seems to agree better with DSSV08 than with BB10. 
\begin{table}[b]
  \centering
  \begin{tabular}{cccccc}
    \toprule
                        & NNPDFpol1.0     & DSSV08~\cite{deFlorian:2009vb}  & BB10~\cite{Blumlein:2010rn}    & LSS10~\cite{Leader:2010rb}   & AAC08~\cite{Hirai:2008aj}          \\
    \midrule
    $\Delta\Sigma(Q^2)$ & $0.31 \pm 0.10$ & $0.25  \pm 0.02$ & $0.19 \pm 0.08$ & $0.21 \pm 0.03$ & $0.24 \pm 0.07$ \\
    $\Delta g(Q^2)$     & $-0.2 \pm 1.4$  & $-0.10 \pm 0.16$ & $0.46 \pm 0.43$ & $0.32 \pm 0.19$ & $0.63 \pm 0.81$ \\
    \bottomrule
  \end{tabular}
  \caption{The first momenta of the singlet and gluon polarised PDFs at the scale $Q^2=\unit{4}{\GeV^2}$ in the 
$\overline{\mbox{MS}}$ scheme. All uncertainties shown are statistical only.}
  \label{Tab:momenta}
\end{table}

Finally, we compute the first momenta of polarised singlet and gluon PDFs 
\begin{equation}
\Delta\Sigma (Q^2) \equiv \int_0^1 dx \Delta\Sigma (x, Q^2) \mbox{ ,}
\ \ \ \ \ \ \ \ \ \
\Delta g (Q^2) \equiv \int_0^1 dx \Delta g (x,Q^2)
 \nonumber
\end{equation}
at the scale $Q^2=\unit{4}{\GeV^2}$ 
and compare with the results obtained by other collaborations (Tab.~\ref{Tab:momenta}, the AAC08~\cite{Hirai:2008aj} 
results are given at $Q^2=\unit{1}{\GeV^2}$). 
Again, we notice the uncertainties of our results: the error on the singlet momentum is between two and four 
times larger than that from other collaborations, while the error on gluon momentum is almost one order of 
magnitude larger. 
\ \\
 
\noindent More precise determinations of polarised PDFs will have to resort to 
data coming from other processes but DIS, 
such as open charm and jet production in fixed target experiments or inclusive jet and $W$ boson production
in proton-proton collisions. 
We plan to extend our analysis to these   data in the near future, and
also to use our PDF set 
to determine the strong coupling constant $\alpha_s$.  

\section*{Acknowledgements}
The research of J.~R.
has been supported by a Marie Curie Intra--European Fellowship
of the European Community's 7th Framework Programme under contract
number PIEF-GA-2010-272515.

 
%
 

{\raggedright
\begin{footnotesize}


\bibliographystyle{DISproc}
\bibliography{nocera_emanueleroberto}
\end{footnotesize}
}


\end{document}